\documentstyle[11pt,paspconf,psfig]{article}

\begin{document}

\title{MHD Turbulence in Star-Forming Clouds} \author{Mordecai-Mark
Mac Low\altaffilmark{1,2}, Ralf Klessen\altaffilmark{3}, Fabian
Heitsch\altaffilmark{2}}

\altaffiltext{1}{Department of Astrophysics,
American Museum of Natural History, Central Park West at 79th Street,
New York, New York, 10024, United States}

\altaffiltext{2}{Max-Planck-Institut f\"ur
Astronomie, K\"onigstuhl 17, D-69117 Heidelberg, Germany}

\altaffiltext{3}{Sterrewacht Leiden, PO Box 9513, 2300 RA Leiden, The
Netherlands}

\begin{abstract}
We review how supersonic turbulence can both prevent and promote the
collapse of molecular clouds into stars.  First we show that decaying
turbulence cannot significantly delay collapse under conditions
typical of molecular clouds, regardless of magnetic field strength so
long as the fields are not supporting the cloud
magnetohydrostatically.  Then we review possible drivers and examine
simulations of driven supersonic and trans Alfv\'enic turbulence,
finally including the effects of self-gravity.  Our preliminary
results show that, although turbulence can support regions against
gravitational collapse, the strong compressions associated with the
required velocities will tend to promote collapse of local
condensations. 
\end{abstract}


\section{Introduction}

A fundamental unanswered question in star formation is why stars do
not form faster than they are currently thought to.  The free-fall
times $t_{\rm ff}$ of molecular clouds with typical densities are 
\begin{equation}
t_{\rm ff} = (3 \pi / 32 G \rho)^{1/2} = (1.2 \times 10^6 \mbox
{ yr})(n/10^3 \mbox{ cm}^{-3})^{-1/2}, \label{ff}
\end{equation}
where $n$ is the number density of the cloud, and I take the mean
molecular mass $\mu = 3.32 \times 10^{-24}$ g.

In contrast to these Myr collapse times, molecular clouds are
commonly thought to be tens of Myr old ({\em e. g.} Blitz \& Shu
1980).  This lifetime is derived from such considerations as their
locations downstream from spiral arms, the ages of stars apparently
associated with them, and their overall frequency in the galaxy.

Either molecular cloud lifetimes are much shorter than commonly
supposed, or they are supported against gravitational collapse by some
mechanism, presumably related to the supersonic random velocities
observed in them.  The first possibility has very recently received an
intriguing examination by Ballesteros-Paredes, Hartmann, \&
V\'azquez-Semadeni (1999); however the remainder of this paper will
concern itself with the second possibility.

Support mechanisms that have been proposed over the years have
included decaying turbulence from the formation of the clouds,
magnetic fields and MHD waves, and continuously driven turbulence.
Each of these raises questions: how can the decay of decaying
turbulence be drawn out over such long periods; can
magnetohydrostatically supported regions collapsing by ambipolar
diffusion reproduce the observations of molecular cloud cores (Nakano
1998); and what could be the energy source for continuously driven
turbulence?

The usual formulation of the problem with maintaining turbulence
arising from initial conditions is that the turbulence is measured to
be strongly supersonic, and shocks are well known to dissipate energy
quickly.  Arons \& Max (1975) were among the first to suggest that
magnetic fields might solve this problem if they were strong enough to
reduce shocks to Alfv\'en waves, since ideal linear Alfv\'en waves
lose energy only to resistive dissipation.  As we will review in more
detail below, Mac Low et al. (1998) showed that a more realistically
computed mix of MHD waves is not nearly so cooperative, a result
confirmed by Stone, Ostriker, \& Gammie (1998).

On the other hand, if the observed motions come from driving, then the
energy source needs to be identified, the amount of energy it is
contributing must be determined, and how to couple the energy source
to the motions of the dense gas must be explained.  Any clues we can
derive from comparison of turbulence simulations to observations are
helpful (see Mac Low \& Ossenkopf 1999 and Rosolowsky et al. 1999 for
recent attempts to do that).

\section{Computations}

In the rest of this paper we will present computations of compressible
turbulence with and without magnetic fields and self-gravity.  For
most of the models we use the astrophysical MHD code ZEUS-3D (Stone \&
Norman 1992a, b; Clarke 1994).  This is a second-order code using Van
Leer (1977) advection that evolves magnetic fields using a constrained
transport technique (Evans \& Hawley 1988) as modified by Hawley \&
Stone (1995), and that resolves shocks using a von Neumann artificial
viscosity.  We also use a smoothed particle hydrodynamics (SPH) code
({\em e. g.} Benz 1990; Monaghan 1992) with a different formulation of
the von Neumann viscosity as a comparison to our hydrodynamical
models.

All of our models are set up in cubes with periodic boundary
conditions and initially uniform density and, in MHD cases, magnetic
field.  We use an isothermal equation of state for the gas, which is a
good approximation for molecular gas between number densities of
$10^2$~cm$^{-3}$ and $10^9$~cm$^{-3}$ typical of molecular clouds.  A
pattern of Gaussian velocity perturbations is then imposed on the gas,
with the spectrum defined in wavenumber space as desired.  Decaying
models are then left to evolve (Mac Low et al. 1998), while driven
models have the same fixed pattern added in every time step with a
varying amplitude computed to ensure a constant rate of kinetic energy
input over time (Mac Low 1999).  Models with self-gravity use an FFT
solver to integrate the Poisson equation.

\section{Decaying Turbulence}

We used models of decaying turbulence to address the question of
whether magnetic fields could significantly decrease the dissipation
rate of supersonic turbulence, as described in Mac Low et al. (1998).
For these models, we set up an initial velocity perturbation spectrum
that was flat in $k$-space and extended from $k = 2$ to $k = 8$.
Although $k^{-2}$ spectra are often used to drive supersonic
turbulence, a spectrum of Gaussian perturbations with this
$k$-dependence is not a good match to a box full of shocks with a
$k^{-2}$ spectrum---in the latter case the dependence is just the
Fourier transform of a step function.

\begin{figure}
\centering{\hspace*{-0.1cm}
\psfig{file=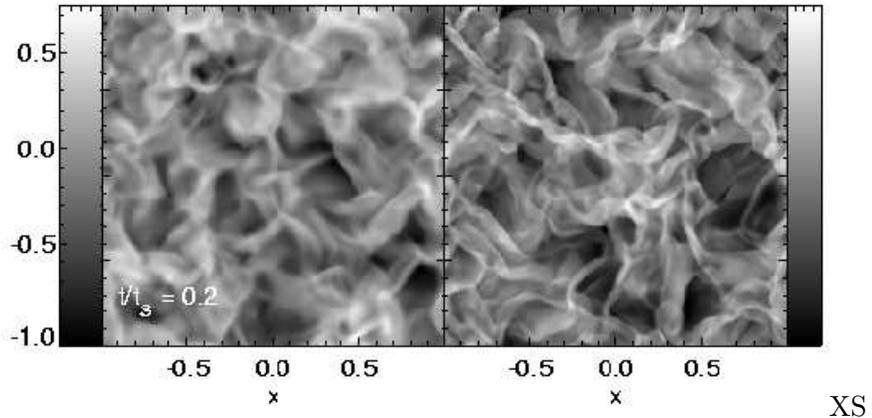,angle=0,width=11cm,clip=}}XS
\caption{Cuts through 3D supersonic turbulence models decaying from an
initial rms velocity of Mach 5, after 0.2 sound crossing times.  The
model on the left was computed with $128^3$ zones, while the model on
the right was computed with $256^3$ zones.  The greyscale is labeled for log
of density \label{decaycut}}
\end{figure}

In Figure~\ref{decaycut} we show examples of cuts through 3D models of
decaying turbulence computed with ZEUS at resolutions of $128^3$ and
$256^3$ (Mac Low 1999).  Although these are cuts rather than column
density images, the tendency for the shock waves to form filamentary
structures reminiscent of molecular clouds can be clearly seen.  

\begin{figure}
\centering{\hspace*{-2in}}
\psfig{file=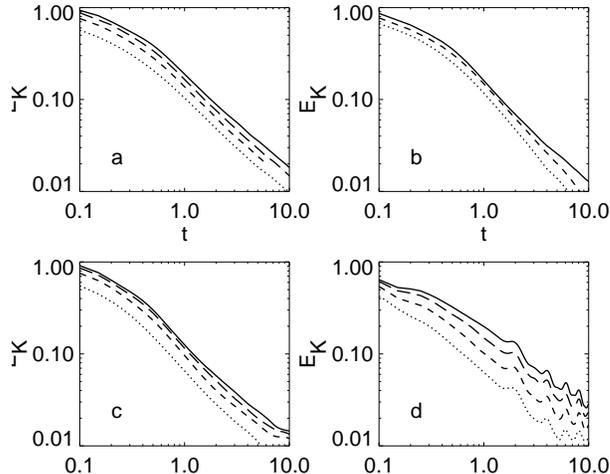,angle=0,width=8cm,clip=}
\caption{Decay rate of kinetic energy over time for 3D computations
with initial Mach number of 5. (a) hydrodyamical with ZEUS (b)
hydrodynamical with SPH (c) MHD with Alfv\'en speed $v_A = c_s$ (d)
MHD with $v_A = 0.2 c_s$.  ZEUS models (a,c,d) were done with $32^3$
(dotted), $64^3$ (short-dashed), $128^3$ (long-dashed) or $256^3$
zones, while SPH models were done with 7000 (dotted), 50 000
(short-dashed) or 350 000 (solid) particles.  From Mac Low et
al.\ (1998). \label{decayrate}} 
\end{figure}

We measured the total kinetic energy on the grid over time for these
models, as shown in Figure~\ref{decayrate}(a).  For comparison, we
also performed a resolution study using SPH, shown in
Figure~\ref{decayrate}(b).  We found that the kinetic energy decays as
$t^{-\eta}$, with $0.85 < \eta < 1.1$ for models in the supersonic
regime (Mach numbers in the range from roughly 1 to 5).  This decay
rate is actually somewhat slower than the decay rate for
incompressible, subsonic turbulence, which, according to the theory of
Kolmogorov (1941), decays with $\eta \sim 5/3$.

We then added initially uniform magnetic fields to see if they could
damp the decay rate.  First we chose a field strong enough for the
thermal sound speed and the Alfv\'en speed to be equal.  As shown in
Figure~\ref{decayrate}(c), the decay rate changed only very slightly,
to $\eta = 0.91$.  Raising the field strength so that the initial
Alfv\'en velocity is unity (Fig.~\ref{decayrate}(d)), we find only
slight further change, to $\eta = 0.87$.  (These results have been
fundamentally confirmed by Stone et al. 1998.)  While this small
decrease in the decay rate is indeed interesting to turbulence
theorists, it by no means fulfills the expectations that magnetic
fields would markedly reduce the energy dissipation from supersonic
random motions.

\section{Driven Turbulence}

\begin{figure}
\hspace*{0.2in}
\psfig{file=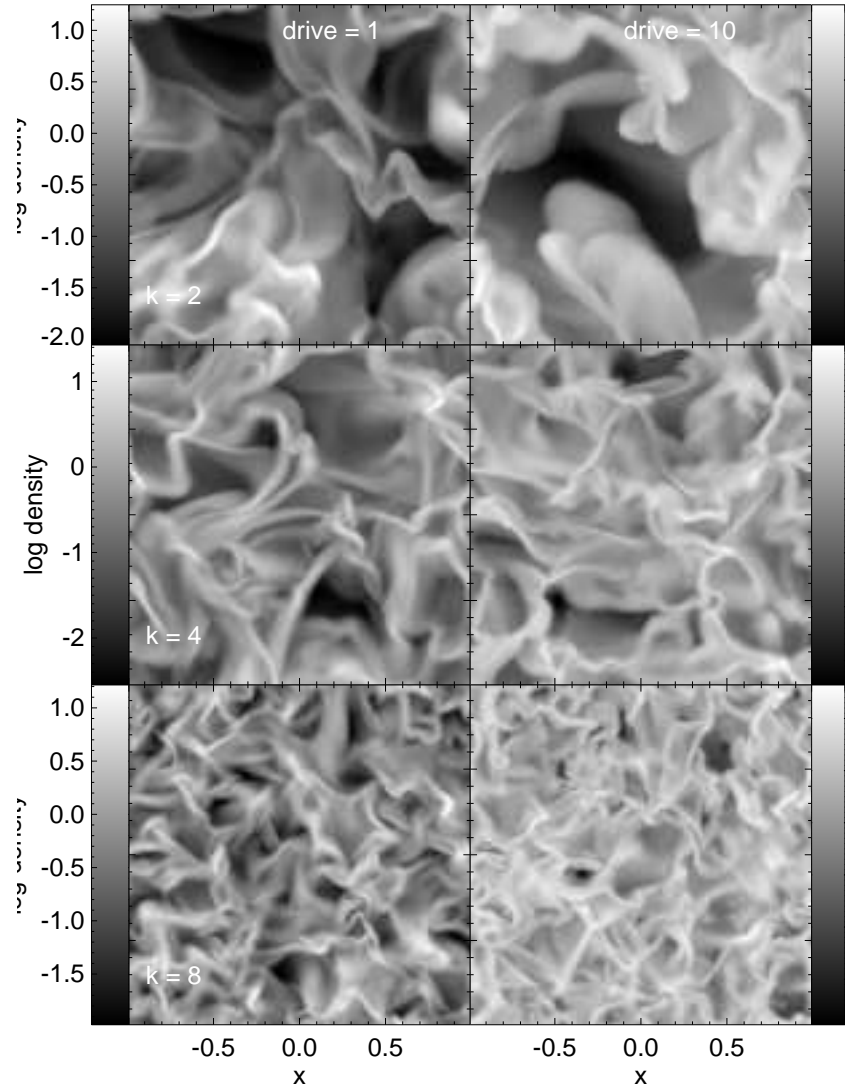,angle=0,width=11 cm,clip=}
\caption{Models showing the appearance of 3D hydrodynamical turbulence
driven with dimensionless wavenumber of $k=2$, 4, and 8, and with
input energies differing by a factor of 10 (``drive'' in the figure.
From Mac Low (1999).  Figures are scaled to their own maximum and
minimum log density. \label{drivencut}}
\end{figure}

In order to try to quantify the decay rate of turbulence, we moved to
models of driven turbulence, as described by Mac Low (1999).  Because
the wavelength of driving strongly influences the behavior of the
turbulence, we used driving functions incorporating only a narrow
range of wavenumbers from $k_D - 1$ to $k_D$, where we only quote the
dimensionless driving wavenumber $k_D$.  In Figure~\ref{drivencut} we show cuts
through two driven models with different wavelengths.

To measure how strongly equilibrium turbulence dissipated energy, we
drove the turbulence with a known, fixed, kinetic energy input rate,
$\dot{E}_K$, and measured the resulting rms velocity $v$.  In the
hydrodynamic case, we found that these quantities excellently followed
the relation
\begin{equation}
\dot{E}_K \simeq (0.21/\pi) m \kappa v^3 \label{ek}
\end{equation}
where $m$ is the mass of the region, and $\kappa = 2\pi/\lambda_D$ is
the dimensionalized wavenumber (for our case, with box-size two,
$\kappa = \pi k_D$), using $\lambda_D$ as the dimensional driving
wavelength.  Although there is some divergence in the MHD case, this
relation is still good to within a factor of two even there.

From this relation, we can compute the decay rate in comparison to the
free-fall time $t_{\rm ff}$ of the region (Mac Low 1999).  If we make the
assumption that $E_K = \int \rho(\vec{x}) \vec{v}(\vec{x}) d\vec{x}
\sim (1/2) m v^2$ (noting that $v$ is the rms velocity in the region),
we can compute a formal decay time $t_d = E_K/\dot{E}_K$ for the
turbulence, by substituting in from equations~\ref{ek} and~\ref{ff} to
find 
\begin{equation}
\frac{t_d}{t_{\rm ff}} = 1.2 \pi \frac{\lambda_D}{\lambda_J}
\frac{1}{M}, 
\end{equation}
where $M = v/c_s$ is the rms Mach number.  Bonnazzola et al. (1987,
1992) have suggested that $\lambda_D < \lambda_J$ is required for
turbulent support to be effective in preventing gravitational
collapse; observations show that $M >> 1$ in typical molecular clouds
(e.g. Blitz 1993), so turbulence appears likely to decay in rather
less than a free-fall time, providing no help to explaining the
apparent long lives of molecular clouds.

The observed random supersonic motions are likely therefore to be
driven.  Four energy sources suggest themselves as possible drivers.
First, differential rotation of the galactic disk (Fleck 1981) is attractive
as it should apply even to clouds without active star formation.
Furthermore, support of clouds against collapse by shear could explain
the observation that smaller dwarf galaxies, with lower shear, have
larger star-formation regions (Hunter 1998).  However, the question
arises whether this large-scale driver can actually couple efficiently
down to molecular cloud scales.  Balbus-Hawley instabilities might
play a role here (Balbus \& Hawley 1998).

Second, turbulence driven by gravitational collapse has the attractive
feature of being universal: there is no need for any additional
outside energy source, as the supporting turbulence is driven by the
collapse process itself.  Unfortunately, it has been shown by Klessen,
Burkert, \& Bate (1998) not to work for gas dynamics in a periodic
domain. The turbulence dissipates on the same time scale as collapse
occurs, without markedly impeding the collapse.  The computations
reported below suggest that magnetic fields do not markedly change
this conclusion.

Third, ionizing radiation (McKee 1989, Bertoldi \& McKee 1997,
V\'azquez-Semadeni, Passot \& Pouquet 1995), winds, and supernovae
from massive stars provide another potential source of energy to
support molecular clouds.   Here the problem may be that they are too
destructive, tending rather to destroy the molecular cloud they act on
rather than merely stirring it up.  If the clouds are coupled to a
larger-scale interstellar turbulence driven by massive stars, however,
perhaps this problem can be avoided.  Ballesteros-Paredes et
al. (1999) even suggest that they are both formed and destroyed on
short time-scales by this turbulence, a possibility well worth further
study. 

A final suspect for the driving mechanism is jets and outflows from
the common low-mass protostars that should naturally form in
any collapsing molecular cloud (McKee 1989, Franco \& Cox 1983, Norman
\& Silk 1980), allowing the attractive possibility of star-formation
being a self-limiting process.  It has recently become clear that
these jets can reach lengths of several parsecs (Bally, Devine, \& Alten
1996), implying total energies of order the stellar accretion energy,
as suggested by Shu et al.\ (1988) on theoretical grounds.  However,
it remains unclear whether space-filling turbulence can be driven by
sticking needles into the molecular clouds.

\section{Self Gravity}

We have begun to investigate directly the support of supersonically
turbulent regions against self-gravity by including self-gravity in
our models of driven turbulence with and without magnetic fields.
Analytic and 2D numerical work by Bonazzola et al. (1987, 1992) and
L\'eorat, Passot, \& Pouquet (1990) suggested that a turbulent Jeans
wavelength could be defined $\lambda_{J,t} \sim \sqrt{v^2/G\rho}$,
where $v$ is again the rms velocity in the region.  They furthermore
specified that the rms velocity differences must be measured at
wavenumbers contained in the region in question.

\begin{figure}
\psfig{file=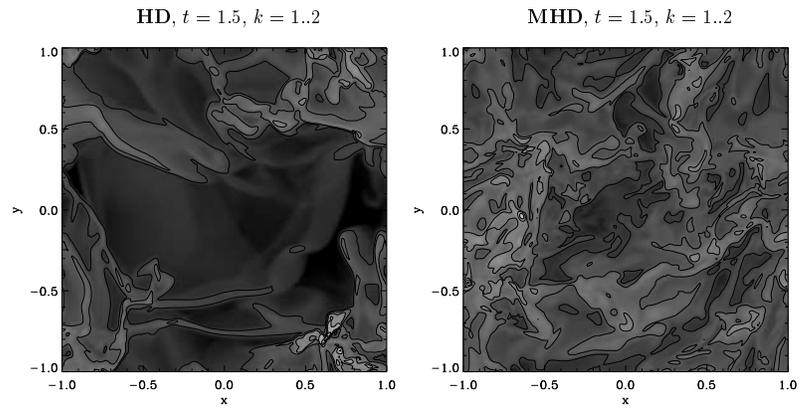,angle=0,width=11 cm,clip=}
\vspace*{-2.7in}
\caption{Comparison of slices of log density for turbulently supported
hydrodynamic and MHD turbulence computed with ZEUS, approximately one
free-fall time after turning on self-gravity.  In both cases, small
regions have begun to collapse, but most of the gas has not (yet) been
accreted by them. \label{collapse}}
\end{figure}

It was already noted by Gammie \& Ostriker (1996) in their 1D MHD
computations that driving could promote collapse as well as preventing
it.  We find that this effect is very significant in our 3D models
where shocks can intersect at multiple angles.  Shocks in isothermal
gas compress the gas by a factor of the square of the Mach number, so
local densities in a region being supported by supersonic turbulence
can exceed the average density by orders of magnitude.  The free-fall
time and Jeans length drop accordingly in these regions, leaving them
no longer supported by the global motions.  In Figure~\ref{collapse}
we show examples of such collapsed regions in turbulence driven with
low and high wavenumbers.  The only way to prevent the collapse would
be to drive the turbulence at such high power and wavenumber that even
regions compressed orders of magnitude above the average were still
supported, which appears astrophysically unlikely (driving wavelengths
would have to be under $10^3$ AU if we take typical molecular cloud
parameters for our models).  Adding magnetic fields with strengths
insufficient to allow magnetostatic support so far appears to make no
qualitative difference to these results.

From our preliminary models it appears that global collapse with high
star-formation efficiency can be at least strongly delayed, if not
prevented, by driven turbulence, but local collapse with low
star-formation efficiency will be forced.  This leads us to speculate
that regions of isolated star formation may correspond to regions
supported by supersonic turbulence, while regions of clustered star
formation may correspond to regions where the turbulence has been
overwhelmed, either by the decay of the local turbulent motions or by
the accretion of additional mass due to large-scale flows (e.g. in
spiral arms or starburst regions).



\acknowledgments Computations discussed here were performed at the
Rechenzentrum Garching of the Max-Planck-Gesellschaft, at the National
Center for Supercomputing Applications, and at the Hayden Planetarium
of the American Museum of Natural History.  I have used the NASA
Astrophysical Data System Abstract Service in the preparation of this
paper.

\end{document}